\input harvmac
\newcount\figno
\figno=0
\def\fig#1#2#3{
\par\begingroup\parindent=0pt\leftskip=1cm\rightskip=1cm\parindent=0pt
\baselineskip=11pt
\global\advance\figno by 1
\midinsert
\epsfxsize=#3
\centerline{\epsfbox{#2}}
\vskip 12pt
{\bf Fig. \the\figno:} #1\par
\endinsert\endgroup\par
}
\def\figlabel#1{\xdef#1{\the\figno}}
\def\encadremath#1{\vbox{\hrule\hbox{\vrule\kern8pt\vbox{\kern8pt
\hbox{$\displaystyle #1$}\kern8pt}
\kern8pt\vrule}\hrule}}

\overfullrule=0pt

\Title{MIT-CTP-2650}
{\vbox{\centerline{ Emission rates, the Correspondence Principle}
\centerline{and the Information Paradox }}}
\smallskip
\smallskip
\centerline{Samir D. Mathur\foot{E-mail: me@ctpdown.mit.edu}}
\smallskip
\centerline{\it Center for Theoretical Physics}
\centerline{\it Massachussetts Institute of Technology}
\centerline{\it Cambridge, MA 02139, USA}
\bigskip

\medskip

\noindent

When we vary the moduli of a compactification it may become 
entropically favourable at some point for a state of branes
and strings to rearrange itself into a new configuration.
We observe that for the elementary string with two large 
charges such a rearrangement happens at the `correspondence point' 
where the string becomes a black hole.
For smaller couplings it is entropically favourable for 
the excitations to be vibrations of the string, while
for larger couplings  the favoured excitations are 
pairs of solitonic 5-branes attached to the string; this helps
resolve some recently noted difficulties with matching emission
properties of the string to emission properties of the black hole.
We also examine the change of state when  a black hole is placed in
a spacetime with an additional compact direction, and the size of 
this direction is varied. These studies suggest a mechanism that 
might help resolve the information paradox.

\Date{June, 1997}

\def\LS{{L^{(S)}}}

\def\TS{{T^{(S)}}}

\def\np {{  Nucl. Phys. }}
\def \pl {{  Phys. Lett. }}

\def \prl {{  Phys. Rev. Lett. }}
\def \pr  {{ Phys. Rev. }}

\newsec{Introduction.}

We can understand many properties of black holes if we use string theory 
as the underlying theory of quantum gravity. A key idea is to compare
 the properties of string theory states at small coupling with
their properties at larger coupling. At small coupling we expect to
 understand the properties of the state from our knowledge of string theory,
while at large coupling the state should behave like a black hole, and should
exhibit the thermodynamic properties associated to black holes
 \ref\suss{L. Susskind, hep-th/9309145;
J. Russo and L. Susskind, {\it Nucl. Phys.}~{\bf B437} (1995) 611.}\ref\sen{
A. Sen,
{\it Nucl. Phys.}~{\bf B440} (1995) 421 and {\it Mod. Phys. Lett.}
~{\bf A10} (1995) 2081.}\ref\stromvafa{ A. Strominger and C. Vafa, 
\pl B379 (1996) 99, hep-th/9601029.}\ref\callanmalda{C.G. Callan and
 J.M. Maldacena, \np B472 (1996) 591, hep-th/9602043.}\ref\dharetal{A. 
Dhar, G. Mandal and S.R. Wadia, Phys. Lett. B388 (1996) 51, hepth 9605234.}\ref\dasmathurone{S.R. Das and S.D. Mathur,
Nucl. Phys. {\bf B478} (1996) 561, hep-th/9606185. }\ref\dasmathurtwo{S.R.
 Das and S.D. Mathur, Nucl. Phys.
 {\bf B482} (1996) 153, hepth 9607149.}\ref\gubserklebanov{S.S. Gubser and 
I.R. Klebanov, 
Nucl. Phys. B482 (1996) 173, hep-th/9608108.}\ref\maldastrom{J.M. Maldacena
 and A. Strominger, Phys. Rev. D55 (1997) 861, 
hep-th/9609026.}\ref\cgkt{C.G. Callan, 
Jr., S.S. Gubser, I.R. Klebanov and A.A. Tseytlin,
hep-th/9610172.}.

\subsec{Extremal and near-extremal states}

The most precise results have been obtained for black holes that are 
extremal or close to extremal. Consider the model of a black
hole in type IIB string theory, in  4+1 noncompact spacetime dimensions.
 Out of
the 10 dimensions of Minkowski space, 5 are compactified on a 5-torus
$T^5=T^4\times S^1$. Let the coordinates $X^5, X^6, X^7, X^8$ span the
 $T^4$ and
the coordinate $X^9$ be along the $S^1$. One way to make a black hole is 
to wrap
$n_5$ D-5-branes on the torus $T^5$, wrap $n_w$ D-strings on circle $S^1$,
 and
consider a momentum $p={2\pi n_p\over L}$ along the direction $S^1$. 
Here $L$ is
the length of the circle $S^1$ and we let the volume of the $T^4$ be
 $V_4$. 

If all the vibrations contributing to the total momentum $p$ travel
 in the same direction along the D-strings, then we have a BPS state, which
would at larger coupling go over to an extremal black hole carrying the
  three
charges corresponding to $n_5, n_w, n_p$. The count of string theory
 microstates
with these charges gives an entropy that agrees with the Bekenstein-Hawking
entropy of the extremal black hole \stromvafa\callanmalda. If we have 
vibrations travelling on the
D-strings in both directions, then we have more energy than the minimum 
required
to carry the charges of the state, and we have a non-BPS string state which
corresponds at larger coupling to a non-extremal black hole. If we are close to
extremality, the extra entropy as a function of the extra energy for the string
state agrees with the corresponding quantity for the non-extremal black hole 
\callanmalda.

The string theory state above can also absorb and emit quanta when the
 vibrations moving in opposite directions on the D-strings collide with each
other. Since the strings are bound to the 5-branes, we assume that they can
vibrate only inside the 5-branes, namely in the directions $X^5, X^6, X^7,
 X^8$.
This leads to the fact that at leading order in the energy, out of the
 10-dimensional gravitons only the $h_{ij}$ with $i,j=5,6,7,8$ can be
 absorbed; in other words scalars of
the 4+1 dimensional spacetime theory can be absorbed but  vector  particles
 and 
gravitons will be suppressed \dasmathurone . This agrees with the expectation
 from the black
hole, which at low energy absorbs scalars but repels higher spin quanta due 
to a
`centrifugal barrier'. The absorption cross section equals the area of the
horizon in the classical calculaton \dharetal\dasmathurone\ref\dgm{S. Das,
 G. Gibbons and S.D. Mathur, 
\prl 78 (1997) 417, hep-th/9609052.}, and it is found that the string theory
state has exactly the same absorption cross section \dasmathurone .

If the momentum charge and amount of nonextremality is small compared to the
 other two charges, the absorption cross  section arising from creation of 
 vibration modes can be
seen to agree with the classical black hole expectation not only at leading
order in energy, but for all energies of the order of the temperature of the
black hole \maldastrom . This gives the agreement of greybody factors between
 the string
theory model and the corresponding black hole. The string model emission rate
 is
proportional to the product of the number of left moving vibrations and the
number of right moving vibrations, and it is interesting that that classical
greybody factor exhibits the same structure. The different numbers of left and
right excitations define different left and right temperatures; the temperature
of the black hole is the harmonic mean of these two temperatures.

\subsec{The correspondence principle, and potential difficulties}

The above results were obtained for black holes that were close to
 extremality. What can we say about more general black holes? A general
conjecture was made by Horowitz and Polchinski about the transition between a
string theory state and the black hole carrying the same charges 
\ref\HP{G. Horowitz and J. Polchinski, hep-th/9612146.}. We
will be interested in the case where the string theory state is just an
elementary string at small coupling $g$, so we discuss the `correspondence
principle' proposed in \HP\ only in this context.

We consider type IIB sring theory with the spacetime  compactified to
 4+1 dimensions as discussed above for the near extremal black hole. Let the
elementary string have winding number $n_w$ and momentum $p={2\pi n_p\over L}$
along $X^9$. If we have a BPS state (all the momentum is carried by the left
movers, there are no right moving vibrations on the string) then, at large $g$,
the metric of this string state would be that of an extremal black hole with
zero horizon area. If we add some extra energy to make the state nonextremal,
then the corresponding metric will describe a black hole with a nonsingular
horizon. Let us fix the Einstein metric of the black hole in the 4+1 dimensional
spacetime, and imagine reducing the coupling $g$ while allowing the string
length to increase. Around some value $g=g_1$ we will find that the curvature of
the string metric becomes of the order of the string scale, and the string
theory description of the spacetime is no longer expected to be a good one. We
will call this point $g=g_1$ as the `correspondence point' or the `matching
point'. What happens for $g<g_1$? The conjecture of \HP\ is that for $g<g_1$ we
can describe the state by using essentially the free elementary string, with
charges $n_w, n_p$ and additional vibrations to make the string energy  equal to
the energy of the  black hole. The agreement of the black hole and free string
descriptions at $g=g_1$ is not expected to be exact, but all physical quantities
of interest should have the same order in either description.

It was shown in \HP\ that when one uses the correspondence principle then the
 entropy of the free string state at the matching point $g=g_1$ agrees with the
entropy of the black hole, upto a  factor of order unity. This provides evidence
in support of the  correspondence principle. But there are some indications that
the matching between black holes and free string states might not be that 
simple:

(i)\quad  The emission rate of low energy quanta from an elementary string
 state is twice that expected from the corresponding black hole 
\ref\HKRS{E.~Halyo, B.~Kol, A.~Rajaraman and L.~Susskind, hep-th/9609075.}.
 This is not a
contradiction in itself, since factors of order unity are not fixed in the
transition between the black hole and the string descriptions. But the factor of
2 can be traced to the fact that the entropy of the string state comes from
vibrations of the string in all 8 directions transverse to the string, rather
than the 4 directions allowed in the near-extremal models discussed above. But
the restriction of the vibrations in the latter case to these 4 compact
directions also implied that only 4+1 dimensional scalars will be emitted at low
energy, while vectors and gravitons will be suppressed. Classically we find that
not just the near extremal hole but also other holes like the Schwarzschild
black hole
 share the property that the low energy emission is dominated by scalars. 
The elementary string state which has vibrations in all 
8 transverse directions seems to emit  vector particles and gravitons with the
same rate as it emits scalars. We need to understand why this should change as
 we enter the black hole phase.

(ii)\quad It was recently argued by Emparan 
\ref\emparan{R. Emparan, hepth 9704204.} that the greybody
 factors suggested by emission by the string state are not the same as those
required by the classical black hole. Again there is no direct conflict with the
correspondence principle, since the at the energies where the greybody
calculation is valid the two different greybody predictions differ by a factor
of order unity, and such factors are not fixed in the correspondence principle.
But the string state has in general unequal number of left and right
excitations, and thus unequal left and right temperatures, and the greybody
factors reflect this fact. But with just two charges, the winding and the
momentum, the classical hole has greybody factors with equal left and right
temperatures. Again we need to understand why this kind of change should occur
as we move from the free string phase to the black hole phase.

\subsec{Summary of results}

In this paper we do the following. We consider the toroidal compactification of
10-dimensional spacetime down to 4+1 dimensions, as described in subsection 1.1.
We take an elementary string with winding charge $n_w$ and momentum charge
$n_p$ in the direction $X^9$. We take $n_w>>1, n_p>>1$. At large coupling $g$
this should give a black hole with two large charges. At small coupling we have
essentially the free string state, and the entropy is carried in left and
right moving vibrations of the string. At the coupling increases past the 
value at the
correspondence point we find that it becomes entropically more advantageous
 to put the
available energy into exciting pairs of solitonic 5-branes, attached to the
elmentary string. Thus as we leave the free string phase (which we call phase
I) and enter the black hole phase (which we call phase II) we have a change in
the way that excitations are carried by the system. 

It was noted in \ref\maldasuss{J.M. Maldacena and L. Susskind, 
\np B475 (1996) 679, hep-th/9604042.} that in the `fat black hole' limit 
the excitations will be in what we have called Phase II, since this kind 
of excitation has the largest entropy when the excitation energy is very
 high. What we are noting here is that the transition from Phase I to 
Phase II happens at the correspondence point of \HP , which is defined 
by the horizon size becoming string scale.

The excitation of  5-branes has the right properties to reproduce the radiation
from the black hole. Let us consider the model, which we call model A, described
by the charges
\eqn\noneone{A:~~~elementary ~string ~winding: n_w, ~~momentum:
n_p,~~solitonic ~5-brane: n_5}
By S-duality we can map this to model B which has the charges
\eqn\noneonep{B:~~~D-string ~winding: n_w, ~~momentum:
n_p,~~D-5-brane: n_5}
By a sequence of S and T dualities we can permute the three charges in model B
in any way we wish  \ref\hms{G. Horowitz,
 J. Maldacena and A. Strominger, Phys. Lett {\bf B383} (1996) 151,
hep-th/9603109.}; in particular we can map to model C with charges:
\eqn\noneonepp{C:~~~D-5-brane: n_w, ~~D-1-brane: n_p,~~momentum: n_5}

The elementary string state we have gives model A with $n_5=0$. This thus has
the same emission properties as model C (with with zero momentum charge).
 Such a model is known to give satisfactory Hawking radiation rates at 
low energy
\dasmathurone\maldastrom .

In the above example we  imagined holding the excitation levels of the
 string fixed while increasing the coupling to reach the change of 
excitation type. We can also consider the case where
a black hole is confined to a compact circle, and then the size of this 
circle is reduced. Following the discussion in \HP\ we expect that there
 should be a change  of excitation type when the circle size becomes 
small enough; we examine this transition.

In outline we perform the following steps:

(a)\quad  We review the correspondence principle of \HP\ for 
the case of an elementary string state. We note that for states with significant
amount of charges the total entropy of the string at the matching point is of
the same order as the entropy of the BPS state with the same charges, which in
turn is of order $\sim \sqrt{n_wn_p}$.

(b)\quad We consider the energy available for non-BPS excitations at the
correspondence point, and find the number  $n_{5\bar 5}$ of solitonic 5-brane
pairs that
can be created with this energy. Note that when there are winding and
momentum charges then the solitonic 5-brane excitations will be fractional \ref\dasmathur{S.R. Das and S.D. Mathur, Phys. Lett. B375 (1996) 103, 
 hep-th/9601152.}\maldasuss , and
it is these fractional pairs that are being counted by $n_{5\bar 5}$. We find
that at the correspondence point the energy which is available to the non-BPS
excitations is always of the order of the mass of one full solitonic
5-brane,
so that the number of fractional excitations is $\sim n_wn_p$, and the entropy
of such excitations is therefore $\sim \sqrt{n_wn_p}$. Thus at the
correspondence point it is equally efficient from the point of view of entropy 
to store the energy in the vibrations of the string or in the creation of
5-brane pairs. For smaller coupling we show that the string vibrations are more
efficient while for larger coupling the 5-brane pairs are more efficient.

(c)\quad We note that in Phase II (where the excitations are the 5-brane pairs)
the emission properties agree with the properties expected of the black hole
carrying two large charges. 

(d)\quad We examine some effects of the gravitational field of the string at
the correspondence point.  In
particular we observe that at the horizon the length of the circle where the
string is wrapped is such that the energy of a free string carrying the given
charges would be minimised. We discuss the relation of this result with
properties of the absorption of higher partial waves by the black hole.

(e)\quad We examine the change in the state of the 4+1 dimensional black hole
 when an additional circle is compactified and the size of this circle is
 reduced. We find that when the size of the compact circle becomes $\sim r_0$ 
 ($r_0$ is the nonextremality length scale of the hole) then it is entropically 
favourable to use the non-BPS energy to excite Kaluza-Klein monopoles wrapping 
around the new compact direction.

(f)\quad We examine how the occurence of fractional charges may lead to a 
change in the picture of how a black hole absorbs an incoming quantum. 
The occurence of a large length scale due to fractional charges may help 
resolve the information retrieval issue for black holes.

(g)\quad  We examine some consequences of a  recent postulate for the quantised 
spectrum of excitations of the winding-momemtum-5-brane system. We observe that
this quantisation appears to describe Phase II but not Phase I.

The plan of the paper is the following. In section 2 we review the arguments of
\emparan\ concerning the disagreements of greybody factors between the elementary
string and the black hole. In section 3 we review the correspondence principle
in our case of interest and demonstrate point (a) above. In section 4 we
demonstrate   point (b).  Section 5 discusses point (c). Section 6
concerns point (d).  Section 7 discusses point (e). Section 8 examines point
 (f). Section 9 is a general discussion. Point (g) is discussed in the Appendix.

\newsec{Emission properties of the elementary string.}

In this section we review the  behavior of emission rates in the context of
the correspondence principle. One issue is that the spins of the quanta emitted
from the elementary string at weak coupling are not those that we expect the
black hole to emit at low energies; this issue was mentioned in the
introduction. We discuss now the details of the greybody spectrum that do not
seem to agree between the black hole and the elementary string; here we will try
to paraphrase some of the arguments of \emparan .

Let us consider the case of the 4+1 dimensional black hole where we have 
two large charges:
\eqn\nthreeone{r_1~>~r_p~>>~r_5,~r_0}
where 
\eqn\nthreeonep{r_1^2\sim G_N^{(5)}M_1\sim (g^2V^{-1}L^{-1}\LS^8)n_wL\TS\sim
 g^2V^{-1}n_w\LS^6 }
In our notation the tension of the elementary string is $\TS=
{1\over 2\pi\alpha'}={2\pi\over \LS^2}$, so that
 $\LS=2\pi\sqrt{\alpha'}$ and under T-duality a 
compact direction of length $A\LS$ goes to a length $A^{-1}\LS$.
We have taken $r_1>r_p$ without loss of generality, since the two charges
 can be interchanged  by a T-duality. For low energy
quanta the greybody factors were computed in \maldastrom . Here low energy
means that \eqn\nthreetwo{\lambda~>>~r_1} Thus
\eqn\nthreethree{\omega\sim \lambda^{-1} <<V_4^{1/2}g^{-1}n_w^{-1/2}\LS^{-3} }

Consider the process where the elementary string absorbs the incoming 
quantum. If we take the coupling to be very weak and thus ignore any 
redshift effects, then
 the change in level of the string is given through
\eqn\twotwt{\delta (M^2)\sim M\delta M=M\omega\sim n_w L\TS \omega\sim 
\TS\delta N_R}
So
\eqn\twotwone{\delta N_R\sim n_wL\omega<<n_w
LV_4^{1/2}g^{-1}n_w^{-1/2}\LS^{-3}\sim n_w^{1/2}LV_4^{1/2}g^{-1}\LS^{-3} }
But the temperature of the right movers is
\eqn\nthreefive{T_R^*\sim N_R^{1/2} }
(Here we are referring to the temperature for the distribution of
 oscillator levels on the world sheet; this is a dimensionless temperature.) To
find $N_R$, note that extra mass over extremality for the classical black hole
is \eqn\twoel{\delta M={(2\pi)^3LV_4r_0^2\over g^2\LS^8} \sim {LV_4\over
g^2\LS^6}}
where in the last step we have set $r_0\sim \LS$ for the correspondence point.
 If
this mass were to be carried in vibrations, the level would be
\eqn\twothir{N_R\sim M_s\delta M\TS^{-1}\sim {n_wL^2V_4\over g^2\LS^6} } 
Note that if we
take $g<<1$, ${L\over \LS}\sim 1,{V_4\over \LS^4}\sim 1$, then we have
$N_R>>n_w>>1$, so that we can use thermodynamic arguments. But from 
\twotwone ,\nthreefive\ 
 \eqn\nthreesix{{\delta N_R\over T_R^*}~<<~1}
 so that we will
see no interesting greybody factors. (The left movers have $\delta N_L=\delta
N_R$, and $T_L^*>T_R$, so they also give $\delta N_L/T_L^*<<1$.)

On the other hand we know that in the present domain of parameters 
the black hole does have  nontrivial greybody factors which come
 from a set of effective left movers and a set of effective right
 movers. Thus we appear to have a disagreement, but we note that one effect 
that
we have ignored is the redshift, which for
 large charges is significant even at the coupling where the string
 turns into a black hole. 

As long as we have $N_R<<N_L$, we will have $T_R^*<<T_L^*$, and the 
greybody factors will reflect these unequal temperatures. On the other
 hand if we have only two charges nonzero, we know from \maldastrom\ 
that the emission from the classical hole is described by a product of 
left and right thermal factors with {\it equal} tempertaures.

 The string can have equal left and right temperatures if $N_R\sim N_L$.
 But note that the difference $N_L-N_R=n_pn_w$ is fixed by the charges. 
So to have $N_R\sim N_L$ we would need to have much larger $N_R, N_L$ than 
those implied by the correspondence principle analysis where the mass of
 the black hole was equated to the mass of the free string.

While it may well be that when we take into account the redshift effects 
the string must have much larger $N_R, N_L$ than that expected from the
 free string analysis, taking these large values will not allow the 
string entropy $S\sim \sqrt{N_R}+\sqrt{N_L}$
to equal the black hole entropy. Thus the greybody factors do not
 agree very well with the correspondence principle, where we equate 
the properties of the black hole to the properties of a string at the
 point where the black hole description ceases to be adequate \emparan .

\newsec{The correspondence principle and non-BPS entropy}

Let us examine the calculation of the correspondence principle 
 for the case that will be of interest to us. The spacetime is $M^5\times
T^5=M^5\times T^4\times S^1$. The string theory state is that of one elementary
string, which can carry winding and momentum charges along the $S^1$ direction. 

The black hole solution corresponding to these charges and some amount of
 nonextremality is given by the following Einstein metric 
$G_E$ and 5-dimensional dilaton $\Phi$:
\eqn\fourtr{ds_5^2=-f^{-2/3}(1-{r_0^2\over r^2})dt^2+f^{1/3}
[(1-{r_0^2\over r^2})^{-1}dr^2+r^2d\Omega^2_3] }
\eqn\fivetr{f=[1+{r_0^2\sinh^2\alpha\over r^2}]
[1+{r_0^2\sinh^2\sigma\over r^2}] }
\eqn\tennewtr{e^{-2\Phi}=(1+{r_0^2\sinh^2\alpha\over r^2})^{1/2}
(1+{r_0^2\sinh^2\sigma\over r^2})^{1/2} }

The 5-d string metric  $G_S=G_Ee^{4\Phi/3}$ is
\eqn\sevennewtr{ds^2_S=-[(1+{r_0^2\sinh^2\alpha\over r^2})
(1+{r_0^2\sinh^2\sigma\over r^2})]^{-1}(1-{r_0^2\over r^2})
dt^2+
[(1-{r_0^2\over r^2})^{-1}dr^2+r^2d\Omega^2_3] }

The curvature at the horizon of the string metric becomes of order the
 string scale when $r_0\sim \LS$. Thus the black hole description is reasonable
if $r_0>>\LS$, but we expect that there is an alternative description in terms of
a string theory state at $r_0<<\LS$.

The mass is
\eqn\qone{M={(2\pi)^3LV_4r_0^2\over 2g^2\LS^8}[\cosh(2\alpha)+\cosh(2\sigma)+1]}
The charges are
\eqn\qtwo{n_w={(2\pi)^2V_4r_0^2\over 2g^2\LS^6}\sinh(2\alpha)}
\eqn\qthree{n_p={(2\pi)^2L^2V_4r_0^2\over 2g^2\LS^8}\sinh(2\sigma) }

The Bekenstein-Hawking entropy of the hole is
\eqn\qfive{S={(2\pi)^4LV_4r_0^3\over g^2\LS^8}\cosh\alpha\cosh\sigma }

The extremal state with the same charges has the mass
\eqn\qfour{M_{ex}={2\pi n_wL\over \LS^2}+{2\pi n_p\over L} }
We can obtain this result by taking the limit $r_0\rightarrow 0$ and
$\alpha, \sigma\rightarrow\infty$ in such a way as to keep the
charges \qtwo , \qthree\ fixed. If we compute the entropy for
the extremal configuration the same way we get $S_{ex}=0$ since in
this limit
\eqn\qsix{S_{ex}=(2\pi)^2{r_0\over \LS}\sqrt{n_wn_p} }
and $r_0\rightarrow 0$ while the other quantities are held fixed.
But we can trust the horizon geometry to give the entropy only for
$r_0>\LS$. If we put \sen\
 \eqn\nnnthreeone{r_0={1\over \sqrt{2}\pi}\LS}
 rather than $r_0=0$ in \qsix\ then we get 
\eqn\qeight{S_{ex}=2\sqrt{2}\pi\sqrt{n_wn_p} }
which agrees with the  entropy of the BPS state of the free string carrying
  charges $n_p, n_w$ (note that the effective central charge for the free 
string  is $12$).

If we equate the mass \qone\ to the mass of a free string state with
winding number $n_w$ and momentum of $n_p$ units, then we get for the
left and right oscillator excitation numbers
\eqn\qnine{N_R=[{(2\pi)^2LV_4r_0^2\over 4g^2\LS^7}]^2
[3+2\{\cosh(2(\alpha-\sigma))+\cosh(2\alpha)+\cosh(2\sigma)\}]}
\eqn\qninep{N_L=[{(2\pi)^2LV_4r_0^2\over 4g^2\LS^7}]^2
[3+2\{\cosh(2(\alpha+\sigma))+\cosh(2\alpha)+\cosh(2\sigma)\}]}

We can take without loss of generality  $\alpha\ge\sigma\ge 0$. 
We take $g<<1$, and the compactification scales to be order string 
scale; the exact scales will drop of our final estimates. We take the
case where we have  two large charges  at the correspondence point 
\eqn\nnntwoone{\alpha>>1, \sigma>>1, ~~~for~~~r_0\sim \LS}
 For covenience of presentation in the calculation
below we also take $\alpha-\sigma>>1$, though this is not essential to the
argument (dropping this restriction just introduces factors of order 
unity in the relations below).

Then the fact that $\alpha>>1, \sigma>>1$ when $r_0\sim \LS$ gives 
using \qtwo\qthree\  that $n_w>>1, n_p>>1$.

From \qnine\qninep\ we find that
\eqn\qten{{N_L\over N_R}\approx e^{2\sigma}>>1}

The entropy of the free string state is
\eqn\qeleven{S_{st}=2\pi\sqrt{2}[\sqrt{N_L}+\sqrt{N_R}] }
The entropy of the extremal string state carrying the same charges
was given in \qeight . The fraction of the entropy that can be
attributed to the non-BPS excitations is measured by
\eqn\qtwelve{{S_{st}-S_{ex}\over S_{ex} }\approx e^{-\sigma}
\approx\sqrt{{N_R\over N_L}}<<1 }

 Thus we see that in the  case at that we have taken (two large 
charges at the correspondence point)
most of the string entropy at the correspondence point is
actually the BPS entropy, which in turn is $\sim \sqrt{n_pn_w}$.

\newsec{The transition from the string to the black hole}

In this section we compare the entropy that can be carried by vibrations of the
string with that which can be carried by excitation of solitonic 5-brane pairs.

\subsec{Entropies of excitations}

The mass available above extremality is, from \qone\ and \qfour\
\eqn\twoel{ M-M_{ex}\equiv\delta M\approx {(2\pi)^3LV_4r_0^2\over
2g^2\LS^8} }
 The mass of a pair of fractional  5-branes is
\eqn\twotw{m_{5\bar 5}={2V_4L2\pi\over \LS^{6}g^{2}n_pn_w}}

We now need to set $r_0\sim \LS$ to be at the correspondence point. For
convenience let us set $r_0=\LS(\sqrt{2}\pi)^{-1}$, which is the value obtained
in \nnnthreeone .  Then the number of fractional 5-brane pairs is
\eqn\nnnone{n_{5\bar 5}= {\delta M\over m_{5\bar 5}}\approx {n_pn_w\over 2}}
The entropy of these pairs is
\eqn\nnonep{S_{5\bar 5}=2\pi[(\sqrt{n_{5\bar 5}}+\sqrt{n_{5\bar 5}})]=4\pi
 \sqrt{{n_pn_w\over
2}}=2\pi\sqrt{2}\sqrt{n_pn_w} }
where we have used that the effective central charge for these excitations
 is $6$.
If we had excited no 5-brane pairs but had put all the energy into vibrations
of the string, the entropy would have been, using \qeight , \qtwelve\
\eqn\nnfourone{S\approx S_{ex}=2\pi\sqrt{2}\sqrt{n_pn_w} }
so that we get the same entropy at the matching point \nnnthreeone\ for 
the two different
ways of carrying the excitations.

Now let us consider the change of the entropy in the two cases when we add a
small extra bit of energy. We hold fixed the coupling $g$, the moduli and the
charges, but have a small increase in $r_0$. The condition that the 
charges are
fixed gives
\eqn\nnfoursix{\delta\alpha=-{\delta r_0\over r_0}\tanh(2\alpha) }
\eqn\nnfoursixp{\delta\sigma=-{\delta r_0\over r_0}\tanh(2\sigma) }

For the case when the excitations are vibrations of the string, we have
\eqn\nnfourseven{\delta S_{st}=
\delta[2\sqrt{2}\pi(\sqrt{N_R}+\sqrt{N_L})]\approx 
\pi\sqrt{2}{\delta N_R\over
\sqrt{N_R}}\approx
{2\sqrt{2}\pi^3LV_4r_0\delta r_0\over g^2\LS^7}e^\alpha }
where we have used that $N_R<<N_L$, and the inequalities \nnntwoone .
For the case where the excitations are 5-brane pairs,
\eqn\nnfoureight{\delta S_{5\bar 5}=\delta[2\pi(\sqrt{n_{5\bar 5}}+
\sqrt{n_{5\bar 5}})]=
2\pi{\delta n_{5\bar 5}\over
\sqrt{n_{5\bar 5}}}={(2\pi)^4LV_4r_0^2\delta r_0\over
4g^2\LS^8}e^{\alpha+\sigma}}

The ratio is
\eqn\nnfourten{{\delta S_{5\bar 5}\over \delta S_{st}}={\sqrt{2}\pi r_0\over
\LS} e^\sigma}
If we set $r_0$ to the value \nnnthreeone\ which we have used for the
 correspondence point
then we get

\eqn\nnfourtenp{{\delta S_{5\bar 5}\over \delta S_{st}}= e^\sigma}
Since $e^\sigma>1$ we see that for $r_0>{\LS\over \sqrt{2}\pi}$
we have $S_{5\bar 5}>S_{st}$ while for $r_0<{\LS\over \sqrt{2}\pi}$
we have $S_{5\bar 5}<S_{st}$.

\subsec{Interpretation}

We have studied above in detail the case of \HP\ that pertains to large winding
and momentum charges. Instead of focusing on the curvature of the metric we
have focused on the microscopically most efficient way to carry the entropy.
The solitonic 5-branes are  heavy when $g$ is small, but it is interesting
that the   values of $g$ and the moduli where they start becoming relevant is
also the set of parameters where the curvilinear metric is starting to be a good
description of the black hole. More generally when we put a string theory state 
in a compact space and change the coupling then at some point the state
 begins to feel the effects of compactification and the excitation 
spectrum changes \ref\mathur{S.D. Mathur, hepth 9609053.}.

By duality we can map the case studied above to model C \noneonepp .  In the
extremal configuration we have $n_w$ D-5-branes and $n_p$ D-strings bound to
these D-5-branes. Clearly the entropy is very small if these D-strings are
joined up to one long string; there will instead be a microcanonical ensemble
of bound states of various winding numbers, and this ensemble has the entropy
$\sim \sqrt{n_pn_w}$.

Now if we add a small amount of nonextremal energy, we do not expect things to
change much. But beyond a certain amount of nonextremality it would be more
advantageous for the D-strings to join up to one long string, so that the
momentum excitations can occur in a fraction    $1/(n_pn_w)$ of one unit of
momentum in the $S^1$ direction. As argued in \maldasuss\ for large excitation
 energies
this will be the favoured mode of excitation; what we note here is that the
changeover occurs exactly at the correspondence point.

[The fact that we have an exact rather than an approximate agreement 
of entropies at the correspondence point \nnonep , \nnfourone\ is not
 a significant fact; this was arranged for convenience by the choice 
\nnnthreeone . We have matched the entropy of the string to the black
 hole entropy in the choice \nnnthreeone\ (this choice actually 
concerned the extremal case \qeight , but the extremal and near 
extremal entropies are very close by \qtwelve ). On the other hand 
we know that the entropy of the near extremal three charge system 
agrees with the entropy of the corresponding black hole \callanmalda . 
Thus we have arranged for the two entropies to agree exactly by the 
choice of the correspondence point.]

\newsec{Emission properties at the correspondence point}

\subsec{Spins of emitted quanta}

Consider our case where we have two large charges at the correspondence
 point.
If we have emission from the free string, then all the 8 directions
 transverse to the string are on equal footing, and so we emit 
5-dimensional scalars, vectors and gravitons at low energy. But
 from our discussion of the above sections at the correspondence
 point where the physics of the string becomes the physics of a
black hole we have instead the low energy excitations as the
 5-brane pairs. By duality we can map this case (which is model A)
    to the model C. The excitations map to momentum and antimomentum 
 modes. But in this latter model we know that we emit at low energy
 5-dimensional scalars $h_{ij}, i,j=5,6,7,8$, while the 5-dimensional
 vectors
$h_{i\mu}, B_{i\mu}$ and the 5-dimensional gravitons $h_{\mu\nu}$
 are suppressed.  Reversing the sequence of dualities, we find that
 the scalars $h_{ij}$ of model C map to the same scalars in model A,
 the $h_{i\mu}$ and 
$B_{i\mu}$ are interchanged, and the $h_{\mu\nu}$ also maps to itself.
 So we see that only 5-dimensional scalars will be emitted at by the
 elementary string once it reaches the correspondence point and passes
 into the black hole phase.

\subsec{Leading order emission rate}

In \HKRS\ it was found that the low energy emission rate from the free
 string was twice what would be expected from the black hole with the
 same charges. But if the excitations that collide and emit quanta are
 the 5-brane pairs then as in the above subsection, the calculation of
 emission rates becomes under duality the collision of momentum modes 
in model C, and here we know that the emission rate does agree with 
the semiclassical calculation of Hawking radiation. Thus the factor 
of 2 found in \HKRS\ will disappear at the correspondence point, at
 least for the elementary string that has large $n_w, n_p$.

\subsec{Greybody factors}

It was argued in \emparan\ that when there are two large charges 
on the elementary string  then we have difficulties matching the
 greybody factors at the correspondence point. The left and right
 temperatures of a free string would be unequal in this situation,
 while the classical cross section demands equal temperatures. But
 if we note that at the correspondence point the non-BPS excitations
 are not the vibrations of the string but the solitonic 5-brane pairs
 then we find that the left and right temperatures for these 
excitations are equal. This can be seen again from the same
 duality as used above. 

\newsec{Gravitational effects}

One of the interesting effects noted in \HP\ was that if we consider 
the gravitational field of the string state at the correspondence
 point, then
the redshift effects will not be small if there are two large charges,
 and this
redshift in fact implies that the asymptotic temperature maps to the 

Hagedorn
temperature at the horizon. Here we note some other effects of the 
gravitational
field of the black hole, in relation to the string theory state which
 gives rise
to the hole.

\subsec{The size of $S^1$ at the horizon}

The 10 dimensional string metric that describes the hole with elementary 
string winding and momentum charges is
\eqn\six{\eqalign{ds^2_S=&[1+{r_0^2\sinh^2\alpha\over r^2}]^{-1}
[-dt^2+(dX^9)^2
+{r_0^2\over r^2}(\cosh\sigma dt+\sinh\sigma dX^9)^2\cr
&+(1+{r_0^2\sinh^2\alpha\over r^2})dX_idX^i]
+
[(1-{r_0^2\over r^2})^{-1}dr^2+r^2d\Omega^2]\cr }}
Consider the extremal limit 
$r_0\rightarrow 0, ~\alpha, \sigma\rightarrow\infty$. In this limit  the
length of the $X^9$ direction at the horizon is
\eqn\nsixone{L_H=L{{\cosh\sigma\over \cosh\alpha}}\rightarrow\sqrt{{n_p\over
n_w}}\LS } Thus this length becomes independent of the length $L$ at infinity.
But we can obtain the same length $L_H$ by the following investigation. Consider
the free string wrapped on the circle, so that there is no effect of gravity and
the metric is flat. Let us ask what is the value of $L'$, the length of the
circle for which the energy of the string is minimised. (We hold fixed the
tension of the string and the charges $n_w, n_p$.) Then we find that we must
minimise \eqn\nsixtwo{M(L')=n_wL'\TS+{2\pi n_p\over L'} } with respect to $L'$,
which gives \eqn\nsixthree{\eqalign{L'_{min}&=\LS \sqrt{{n_p\over n_w}},
~~n_wL'_{min}\TS={2\pi\over \LS}\sqrt{{n_p n_w}},\cr
&~~ {2\pi n_p\over
L'_{min}}={2\pi\over \LS}\sqrt{{n_p n_w}}, ~~ M_{min}={4\pi\over
\LS}\sqrt{{n_p n_w}}\cr} } 

Thus we get $L'_{min}=L_H$, so that the circle size at the horizon
 is such that in a free theory it would minimise the mass for the given charges.
We also note that for this special length the free string state has equal mass
contributions from the winding and momentum charges. 

\subsec{A comment on the absorption of angular momentum}

In \ref\maldastromtwo{J. Maldacena and A. Strominger, hepth 
9702015.}\ref\mathurtwo{S.D. Mathur, hepth 9704156.}\ref\gubser{S. 
Gubser, hepth 9704195.}  the absorption 
of angular momentum by black holes was studied. In \mathurtwo\gubser\ 
it was noted that if an effective string model was to be used for the 
absorption, then the tension of this string would have to be $\sim 
(r_1r_5)^{-1}$ since the classical cross section is a function of 
the product $r_1r_5$. If we take the absorbing element to be  a 
D-string with its naive tension then the tension would be $\sim r_5^{-2}$. 

One possibility is that the details of the bound state of D-1-branes
 and D-5-branes at weak coupling is such that the requisite tension 
is effectively produced at low energies. Here we consider another possibility
for the source of a tension that is symmetric in $r_1$ and $r_5$.

From the analysis of the above subsection we note that if we have
 two large charges, then the near the horizon  geometry is such 
that if we placed
the charges here then they would have equal contributions to the 
local mass. 
When we have a low energy wave incident on a black hole, the wave 
is oscillatory
at infinity, essentially non-oscillatory over the scales $r_1, r_5$ 
and then
oscillatory in the near horizon region due to the increasing blueshift.
 In the calculation of the leading order absorption cross section the 
tension of the effective string drops out \dasmathurone , but it may 
be that for subleading effects we need to use an effective tension that
 includes effects of gravity and uses the near horizon geometry for 
the  effective string analysis of
absorption. In that case the above discussion suggests a reason why
 $r_1$ and $r_5$
enter in a simple symmetric combination in the absorption cross section,
 since now these would correspond the two large charges.

\subsec{The non-BPS mass blueshifted to the horizon}

Consider the system with large charges $n_w, n_p$, with a small amount of
non-BPS excitation, which brings the system to the vicinity of the
correspondence point.  The mass above extremality is then 
\eqn\tenone{\delta
M\approx{(2\pi)^3LV_4r_0^2\over 2g^2\LS^8}} If we consider this mass 
blueshifted
to the horizon, then we would need to multiply \tenone\ by the factor
\eqn\tentwo{ \nu\cosh\alpha\cosh\sigma\approx\nu {g^2\LS^7\over
LV_4r_0^2(2\pi)^2} \sqrt{n_wn_p}}
where following \HP\ we have replaced  $(1-{r_0^2\over r^2})^{-1/2}$ by a
quantity $\nu\sim 1$.
The  mass \tenone\ blueshifted to the horizon is
\eqn\tenthree{\delta M_H= \sqrt{n_wn_p}{\pi\over \LS} \nu}

We observe that this quantity is of the same order as the mass 
$M_{min}$ \nsixthree\ of the BPS
state of the elementary string wrapped at the horizon and carrying 
the charges
of the hole. The interpretation of this coincidence is not clear.

\newsec{The `crushing' transition}

Suppose we have a black hole in $D$ space-time dimensions, and we
 compactify one additional direction on a circle. It was noted in 
 \HP\ that from the viewpoint of classical geometry if the size of
 this circle is much larger than the horizon then we essentially get
 a $D$-dimensional hole, while if the circle is smaller than horizon 
size then  we expect the stable solution to be $D-1$ dimensional hole.
 This happens because for small compactification radius the latter 
solution gives larger horizon area for the  same mass, and is thus 
expected to be the stable solution. In the compactification of branes
 on a circle, it was argued in \HP\ that these two geometries had 
microscopic explanations, in terms of `unwrapped' and `wrapped' branes
 respectively.

We wish to analyse from a microscopic viewpoint this kind of transition 
for the case where we have a black hole in 4 spacetime dimensions, 
and a fifth direction is compactified and taken to be large or small.

Let us start with the 4-dimensional hole.  We let the black hole have 
three charges, corresponding to the charges of our model A. The horizon
 is nonsigular because there is a small amount of nonextremal energy. 
When an addtional direction is compactified in model A, on a circle of
 length $L'$, the extra kind of excitation that is available is pairs 
of  Kaluza-Klein monopoles \ref\halyo{M. Cvetic and D. Youm, Phys. Rev.
 D53 (1996) 584, M. Cvetic and A. Tseytlin, Phys. Lett. B366 (1996) 95,
 J. Maldacena and A. Strominger, hepth 9603060, C. Johnson, R, Khuri 
and R. Myers, hepth 9603061, E. Halyo, hepth 9611175.}. 

The mass above extremality for the 4-dimensional hole is
\eqn\twone{M-M_{ex}\equiv \delta M={(2\pi)^2LL'V_4r_0\over 2g^2\LS^8} }
The mass of a pair of monopoles is
\eqn\twtwo{m_{m\bar m}={2(2\pi)L{L'}^2V_4\over g^2\LS^8}}
Thus the number of  pairs of monopoles that can be created by the mass
 \twone\ is
\eqn\twthree{f={\delta M\over m_{m\bar m}}=
{\pi r_0\over 2L'}}

First let us take the case where the energy above extremality is used
 to create excitations of the three charge system (i.e. there is no 
excitations of the monopoles). The analogue of \qtwelve\ says that 
the entropy is essentially the extremal one, 
\eqn\twfour{S_3\approx 2\pi\sqrt{n_1n_2n_3} }

Now take the case that the non-BPS energy goes to creating the 
monopole-antimonopole pairs. The entropy of this four charge system is
\eqn\twfive{S_4=4\pi\sqrt{n_1n_2n_3f} }
We have used in both cases the fact that $c=6$ 
\ref\klebanovtseytlin{I. Klebanov and A. Tseytlin, hepth 9604179, 9607107.}

The entropies $S_3$, $S_4$ agree when 
\eqn\twsix{2\sqrt{f}=1, ~~f={1\over 4}, ~~r_0={L'\over 2\pi} }

Following arguments similar to those in section 4, we can show that
 the entropy of the kind $S_4$ dominates when $r_0>>{L'\over 2\pi}$,
 while the entropy $S_3$ dominates when $r_0<<{L'\over 2\pi}$. Thus
 the excitations that `see' the extra compactified direction come 
into play just when the size of this direction equals the scale $r_0$. 

We get a similar result if we start with the metric of the 5-dimensional hole
and take one extra compactified direction whose length $L'$ we vary.
The mass above extremality is
\eqn\twoelpn{ M-M_{ex}\equiv\delta M\approx {(2\pi)^3LV_4r_0^2\over
2g^2\LS^8} }
Thus the number of created monopole pairs is
\eqn\twthreepn{f={\delta M\over m_{m\bar m}}=
{(2\pi)^2 r_0^2\over 4{L'}^2}}
Again we obtain equality of $S_3$, $S_4$ when
\eqn\nonen{f={1\over 4}, ~~r_0^2={{L'}^2\over (2\pi)^2}, ~~
r_0={L'\over 2\pi} }

To interpret the scale $r_0$ in say \nonen\ we note that the three kinds
 of charges are symmetric under U-duality. The Einstein metric is
\eqn\fourtrn{ds_5^2=-f^{-2/3}(1-{r_0^2\over r^2})dt^2+f^{1/3}
[(1-{r_0^2\over r^2})^{-1}dr^2+r^2d\Omega^2] }
\eqn\fivetrn{f=[1+{r_0^2\sinh^2\alpha\over r^2}]
[1+{r_0^2\sinh^2\sigma\over r^2}][1+{r_0^2\sinh^2\gamma\over r^2}] }
In analogy to \tennewtr\ we define
\eqn\tennewtrn{e^{-2\tilde \Phi}=(1+{r_0^2\sinh^2\alpha\over r^2})^{1/2}(1+{r_0^2\sinh^2\sigma\over r^2})^{1/2} 
(1+{r_0^2\sinh^2\gamma\over r^2})^{1/2}}

Then we define the metric
$\tilde G=G_Ee^{4\tilde \Phi/3}$ 
\eqn\sevennewtrn{d\tilde s^2=-f^{-1}(1-{r_0^2\over r^2})
dt^2+
[(1-{r_0^2\over r^2})^{-1}dr^2+r^2d\Omega^2] }
In this metric $r_0$ is the size of the horizon.

\newsec{A conjecture on the absorption process}

In a semiclassical picture a quantum infalling into a black 
hole falls smoothly through the horizon into the interior 
of the hole, thus trapping itself causally from the outside
 world to which it can send no information. The Hawking 
radiation that takes away its energy arises from redsfiting
 of the vacuum modes near the horizon, so that an impure quantum
 state is forced to result at the end of the evaporation process. 

If this fate is to be avoided by the string theory black hole then 
it does not appear to be enough that there be a suitable theory of
 quantum gravity at the planck scale; the above argument does not 
get invalidated by the presence of small scale local fluctuations of the
 spacetime \ref\eskomathur{E. Keski-Vakkuri and S.D. Mathur,
 \pr D54 (1996) 7391.}. Nor does it help that the string scale
 may be somewhat longer than the planck length, since the black
 hole horizon scale can be taken as big as we wish. What we would
 like  is a length scale that grows with the size of the hole. 
Such a length scale can arise from the property of fractionation
 of branes
\dasmathur\maldasuss , and we give a schematic model that invokes
 this physics below. (It was  pointed out in \maldasuss\ that
 fractionation gives rise to long strings, but the physics and scales
involved there do not appear to be the same as those that will
 arise in our analysis.)

Let us consider absorption into the extremal 4+1 dimensional
 hole for convenience. Our basic postulate will be the following.
 When an incoming quantum comes at a distance $L'$ from the hole,
 then in some sense the situation is like the one where we 
compactify an additional direction with length $L'$. If we 
had compactified another direction then we could have excited 
pairs of (fractional) monopoles-antimonopoles wrapping around 
this new circle; this is what we used in the last section. 
We assume that the incoming quantum can also act as a `peg' 
around which the new kind of excitation can wrap. We do not 
know how to justify this assumption in any rigorous way.

Because of fractionation, the monopole excitations can be quite
 light; in fact we will see that an adequate excitation can arise
just from the kinetic energy of confining the quantum to within a
 horizon radius. If this happens, then we do not have the picture 
of a quantum freely falling through the horizon; instead the structure
 of the hole rearranges itself somewhat and  monopole pairs emerge 
to wrap around the quantum. With such an `active' mode of absorption 
it is plausible that the information of the quantum can be transmitted
 to the emerging radition. 

\subsec{Outline of calculation}

(a)\quad The extremal hole itself has no energy available to create
 the non-BPS monopole pairs. Thus this energy must be supplied by
the incoming quantum; let the energy used be $\delta M$. This energy
 can create some number $f$ of monopole pairs; we expect to get a 
fractional number of pairs, $f<<1$.

(b)\quad In the calcuations of the earlier sections we have taken 
the excitation type to be of entirely of one kind or entirely of 
another kind, and then made a comparison of entropies. While this
 if fine for locating the rough transition point between configurations,
 in the present case we expect that there will be only a small change of 
excitation type when a small quantum arrives. Thus assume for 
convenience that we are in model A and that the excitations of 
the extremal hole are given by counting the fractional momentum
 modes that run of a string of length $n_5n_wL$. Upon arrival of
 the quantum let a fraction $\mu$ of these modes still contribute
 to the entropy in this form, while a fraction $(1-\mu)<<1$ of the
 momentum modes bind to one state (thereby losing entropy) but giving 
rise to monopole pairs that occur in units of $[n_5n_wn_p(1-\mu)]^{-1}$
 of a full pair (thereby increasing entropy). We do not know $\mu$ a
 priori, but we extremise the entropy over $\mu$ and find what the
 best arrangement of excitations would be.

(c)\quad We require that the increase in entropy which occurs upon
 rearrangement of excitations be such that the entropy increase by 
at least order unity. This would indicate that the postulated process
 is dynamically probable, and not just energetically possible.  Note 
that the entropy increase of order unity means that we double the 
available states; it is not enough to ask that that states go from
 a large number $N$ to $N+1$ since in that case there  is a very 
small likelihood of reaching the new excitation within a dynamical
 time scale of the system,

(d)\quad The entropy increase depends on the available energy, 
so we ask what value of $\delta M$ would produce the order unity
 increase in entropy. We find that the required value of $\delta M$
 is $\sim R_H^{-1}$, where $R_H$ is the radius of the horizon. 
If we try to confine a particle trajectory to make it enter the hole,
 then we expect this to be the minimum energy that would accompany 
the particle. Thus the absorption appears allowed by such a mechanism
 for all infalling quanta.

\subsec{Calculations}

 Let the supplied energy be $\delta M$. Let this be enough 
to create a fraction $f$ of a complete monopole  pair. Thus
\eqn\thone{f={\delta M\over m_{m\bar m} } }
Let a fraction $\mu<1$ of the  quanta of the $n_3$ charge be
 distributed in the manner  required to maximise the entropy 
of the three charge system, and let the remainder $1-\mu$ be
 bound up into one state, thus allowing a fractionation of
 monopoles by the factor $n_1n_2n_3(1-\mu)$.
The total entropy of this set of states is
\eqn\thtwo{2\pi\sqrt{n_1n_2n_3\mu}+4\pi\sqrt{n_1n_2n_3(1-\mu) f} }
Let us extremise this with respect to $\mu$. Then we get
\eqn\ththree{{1\over 2\sqrt{\mu} }-{2\sqrt{f}\over 2\sqrt{(1-\mu)}}=0}
\eqn\ththreep{f={1-\mu\over 4\mu}, ~~\mu={1\over 1+4f},~~~
1-\mu={4f\over 1+4f}\approx 4f~~~for ~f<<1 }

[Note that if $f>>1$, then we have $\mu<<1$ and we are in `Phase II' 
where the excitations are monopole pairs. If $f<<1$ then $1-\mu<<1$ 
and we are in `Phase I' where most of the entropy comes from the 
distribution that gives the BPS entropy of the three charge system.]

We will be interested in the case $f<<1$. In that case, 
the entropy gain by taking $\mu$ to be its optimal value, rather than unity, is
\eqn\thfour{2\pi\sqrt{n_1n_2n_3}[(1+4f)^{-1/2}+2\sqrt{({4f\over 1+4f})f}-1]
\approx 2\pi\sqrt{n_1n_2n_3}[-2f+4f]=4\pi\sqrt{n_1n_2n_3}f }

We would like this extra entropy to be order unity. So we have
\eqn\thfive{4\pi\sqrt{n_1n_2n_3}f=1, ~~~f={1\over 4\pi\sqrt{n_1n_2n_3} } }
Thus we need
\eqn\thsix{{\delta M\over m_{m\bar m} }={1\over 4\pi\sqrt{n_1n_2n_3} } ,
 ~~\delta M={m_{m\bar m}\over 4\pi\sqrt{n_1n_2n_3} } }

Taking $m_{m\bar m}$ from \twtwo ,
\eqn\thseven{\delta M={2(2\pi)L{L'}^2V_4\over g^2\LS^84\pi\sqrt{n_1n_2n_3}} }

Note that 
\eqn\theight{2\pi\sqrt{n_1n_2n_3}=A/G^{(5)}_N=
{2\pi^2R_H^332\pi^2LV_4\over g^2\LS^8} }
Thus 
\eqn\thnine{\delta M={{L'}^2\over 32\pi^3 R_H^3} }
If we put $L'= R_H$ (the quantum is in the range of the horizon)
 then we get
\eqn\thten{\delta M = {1\over 32\pi^3 R_H} }
But $\sim R_H^{-1}$ is the minimum energy that will accompany the
 quantum localised within a distance of order the horizon size. 
Thus when the incoming quantum is of the order of the horizon 
distance away then we can create fractional monopole pairs using
 its energy, such that the entropy gain by creating these pairs 
is order unity, and the process is thus seen to be probable and not
 just possible. 

\subsec{Notes on the above calculation}

(a)\quad We have allowed the available non-BPS energy to
 form monopoles pairs that wrap around the incoming quantum,
 but we have not allowed this energy to be used to excite the
 non-BPS excitations of the three charge system itself. The 
latter excitations, which are just the excitations of the right moving
momentum modes in the above example, would in fact have a higher
 entropy than the monopole pairs. But we can imagine that the 
incoming quantum cannot transfer its energy to these momentum 
modes directly, while it can transfer it to the monopole pairs
 since these pairs are the ones that see the location of the quantum.
 After the quantum has been absorbed, the energy can be transferred to
 the right moving momentum modes, which would be entropically more favourable, 
and would also be in accord with the effective absorption process at
 weak coupling 
\dasmathurone .

(b)\quad We have used the formulae for the entropy of fractional 
excitations in a domain where a very small fraction for more than
 one charge is present (eq. \thtwo ). This issue may need a more 
careful analysis.

(c)\quad The smallest energy quantum that can be absorbed by the
 extremal hole has an energy much lower than $\sim R_H^{-1}$ 
\ref\trivedi{J. Preskill, P. Schwarz, A. Shapere, S. Trivedi 
and F. Wilczek, Mod. Phys. Lett. A6 (1991) 2353, C. Holzhey
 and F. Wilczek, \np B380 (1992) 447, P. Kraus and F. Wilczek,
 \np B433 (1995) 403.}\maldasuss . But we have considered the 
infall of a well defined trajectory rather than the absorption
 of a monochromatic wave, and here it seems more reasonable to
 use the scale $R_H^{-1}$ as the minimum energy that the quantum
 must have to fall in. The geometric picture that we have tried 
to make of the absorption process pertains to such localised 
trajectories.

(d)\quad The most unclear step of course is the argument 
that
the infalling quantum sets a scale which can be taken as 
a compactification scale for the generation of pairs of 
the fourth charge. The location of the quantum will change
 with time, and when it enters deep within the hole then we
 expect that the energy has been converted to the right moving vibrations.

\newsec{Discussion}

In this paper we have considered the case of the string state with large
 momentum and winding charges. The transition from the black hole to the
 weakly coupled string, which happens when the horizon is string scale \HP ,
  has also a simple microscopic description. This transition point is 
characterised by parameters and a degree of nonextremality such that for
 smaller energies (or weaker coupling) it is entropically more advantageous 
to store the non-BPS energy in the form of left and right moving vibrations
 of the string, while for larger excitation energies (or larger coupling) 
it is entropically better to unify all the winding and momentum modes to
 one bound state, and to excite fractional piars of solitonic 5-branes
 to carry the non-BPS energy.

As pointed out in \HP\  the rate of growth of entropy as a function of
 the energy is different for  the free string and for the black hole; 
thus agreement can be obtained  only at the `correspondence' point. 
The black hole entropy in the near extremal case is known to agree 
with the entropy of the three-charge system, so it is not a surprise
  that the point where   the non-BPS excitations  will change from 
being string vibrations to being solitonic-5-brane
 pairs will also be the correspondence point.

But with this microscopic picture, we see that there is no reason for
 the properties of emission (spins, greybody factors etc.) to agree 
between the string phase and the black hole phase. In fact for large
 charges, we may term the change of excitation type at the correspondence
 point as a phase transition, since the degrees of freedom that which 
are manifested  undergo a change. While we do not undestand the physics 
of emission from the strongly coupled black hole phase it is gratifying
 that the three charge model which does reproduce the right low energy 
emission at weak coupling
is entropically favoured from the correspondence point onwards into
 the black hole phase.

It has   been recently noted that emission rates fail to agree 
in significant ways for black branes at the correspondence point
 \ref\dastwo{S.R. Das, hepth 9705165.}. It would be worth investigating
 if in this case too there is a change of excitation type that occurs 
at the correspondence point (for example the excitation of a pair of 
higher dimensional branes).

We have also investigated the change in state of a black hole when
 an additional direction is compactified and made smaller so that 
the hole is `crushed'. In accordance with the expectation in \HP\ 
there is a change of the entropically favoured state at a certain
 radius of compactification; we find this radius.

With regard to the information paradox, we note that two places where
 strings differ from usual quantum gravity plus matter theories are (i)
 the   fractionation of quanta by other quanta  \dasmathur\maldasuss\ 
which gives rise to new scales depending of the number of particles 
present and (ii) the occurence of a $U(N)$ gauge theory when $N$ quanta
 of the same type come close together
\ref\witten{E. Witten, \np {B460} (1996) 335.} which encourages the quanta
 to spread out from each other. We have speculated on a mechanism 
using fractionation in this paper, suggesting a more dynamical black 
hole absorption process than a simple infall into a smooth geometry.
Some arguments for the existence of a long length scale using the 
enhanced gauge symmetry were given in
 \ref\klebanovmathur{I. Klebanov and S.D. Mathur, hepth 9701187.};
 these two effects may be closely related in the black hole problem. 

\bigskip
\bigskip
\bigskip

\centerline{{\bf Acknowledgements}}

I am grateful to S.R. Das for discussions and a  critical reading of the
 manuscript.
This work is partially supported by cooperative
agreement number DE-FC02-94ER40818.
 
\vfill
\eject

\appendix{A}{Quantisation of the three-charge system}

To be able to fully quantify the transition between Phase I and Phase II we
need a quantisation of the winding-momentum-5-brane system
analogous to the quantisation of the elementary string. A proposal in this 
direction was given in \ref\larsen{F. Larsen, hepth 9702153.}. 
In this Appendix we (i) perform algebraic manipulations to obtain 
a mass formula in terms of excitation level, the charges and the 
moduli (ii) verify that for a unit increase in excitation number
 the mass increase in a certain limit is what would be expected 
by analogy with the elementary string (iii) observe that the left
 and right temperatures in this quantisation agree with the 
expectation of what we have called Phase II but not with Phase I.

We will assume that we are in model C, though by duality the same
 results hold for any model.

For the elementary
string the mass formula is 
\eqn\nfiveone{m^2=(n_wL\TS+{2\pi n_p\over
L})^2+8\pi\TS N_R }
For later reference we write
\eqn\nnfiveone{Q_w=n_wL\TS, ~~~Q_p={2\pi n_p\over L}}
where the quantities $Q_w, Q_p$ give the mass of the BPS state with
 the given
winding and momentum charge respectively.
 
The neutral excitations are measured by one integer $\delta
N_R=\delta N_L$, and not by different numbers that correspond to winding and
momentum excitations. But if the winding is the dominant contribution to 
the mass
in \nfiveone\ then the excitations have a spectrum that looks like the spectrum
of momentum modes on a single long string \dasmathur :
\eqn\nfivetwo{\delta m\approx {8\pi\TS\over 2m}\approx {4\pi\TS\over n_w
L\TS}={4\pi\over n_wL} }

Now we consider the quantisation of \larsen\ of the three charge system 
but for simplicity   restrict ourselves to the case of no angular momentum.
 Let the mass be
written as 
\eqn\thone{M={\mu\over 2}\sum_i\cosh(2\delta_i) }
Define the effecive charges
\eqn\thtwo{Q_i={\mu\over 2}\sinh(2\delta_i) }
The entropy is then
\eqn\thth{S=2\pi\mu^{3/2}\prod_i\cosh(\delta_i) }
 Then the system is described by left and right oscillator numbers, with
\eqn\thfour{N_R={\mu^3\over 4}[\prod_i\cosh(\delta_i)-\prod_i
\sinh(\delta_i)]^2 }
\eqn\thfourp{N_L={\mu^3\over
4}[\prod_i\cosh(\delta_i)+\prod_i\sinh(\delta_i)]^2} 

In the notation used in \ref\cvetic{M. Cvetic, hepth 9701152.}\larsen\ 
and adopted in this appendix
\eqn\nnfivefive{\prod_iQ_i=\prod_in_i}
which is equal to setting to unity the following quantity of units
 $(length)^3$:
\eqn\nnfivesix{{g^2\LS^8\over (2\pi)^3LV_4}=1 }

\subsec{The mass spectrum}

With some manipulations we can write
\eqn\thfive{M={1\over 2}\sum_i[\mu^2+4Q_i^2]^{1/2} }
\eqn\thsix{\eqalign{N_R=&{1\over 32}\prod_i[(\mu^2+4Q_i^2)^{1/2}+\mu]
+{1\over 32}\prod_i[(\mu^2+4Q_i^2)^{1/2}-\mu]-2\prod_iQ_i \cr
&={1\over 16}\prod_i[\mu^2+4Q_i^2]^{1/2}
+{\mu^2\over 8}M-2\prod_iQ_i \cr }}

We then get
\eqn\thel{[N_R-{\mu^2\over 8}M+2\prod_iQ_i]^2=
{1\over 256}\prod_i(\mu^2+4Q_i^2) }

This is a cubic in $\mu^2$, so we can solve it explicitly for $\mu^2$ as a
function of $M, N_R$, and the charges $Q_i$. Substituting in \thfive\ we get a
relation $f(M,Q_i,N_R)=0$ which is analogous to \nfivetwo .

\subsec{A check on  the mass spectrum}

Consider the case 
\eqn\nnfivetwo{\mu<<Q_3<<Q_1, Q_2}
For concreteness let $Q_3$ correspond to D-5-branes, $Q_1$ to D-strings
and $Q_2$ to momentum. Following what we saw in \nfivetwo\ we wish to see that
if we make a unit change in $N_R$ then the change in mass of the soliton 
complex
should approach the mass of a fractional 5-brane pair, if the winding and
momentum charges are large. Note that this pair should consist of fractional
branes, due to the presence of the other two charges.

In \thsix , let $\delta N_R=1$, $\delta Q_i=0$. Then with \nnfivetwo ,
\eqn\nnfivethree{\mu\delta\mu\approx{8Q_3\over Q_1Q_2} }
\eqn\nnfivefour{\delta M\approx{1\over 4Q_3}\mu\delta\mu ={2\over Q_1Q_2}}
Using \nnfivesix\ we convert this mass change to the notation used elsewhere 
in
this paper:
\eqn\nnfiveseven{\delta M=2{2\pi LV_4\over g\LS^6 n_wn_p }}
which is seen to be exactly the mass of a fractional D-5-brane pair. Thus we
have recovered the analogue of \nfivetwo , which provides one consistency
 check of the
quantisation.

\subsec{The domain of applicability of the quantisation}

In this quantisation the left and right temperatures are \larsen\
\eqn\nnfiveeight{T^{-1}_{R,L}=\pi\mu^{1/2}
[\prod_i\cosh\delta_i\pm\prod_i\sinh\delta_i]}
Thus if one charge (say $Q_3$) is zero, then
\eqn\nnfivenine{T^{-1}_R=T^{-1}_L=\pi\mu^{1/2}\cosh\delta_1\cosh\delta_2}
This equality of temperatures is expected of excitations in Phase II
in out language, but in Phase I the left and right temperatures are not equal.
Thus we see that the quantisation proposal of \larsen\ covers Phase II
 but not Phase I.
So it appears that the proposal cannot be used as a rigorous quantisation
 of the
complete three-charge system. 

The reason that the proposal naturally covers Phase II is straightforward:
 it
was derived from a study of black hole properties which as we have seen
 pertain
to Phase II. The classical hole is described by the limit of large charges.
 In
this situation the dominant contribution comes from the effect of fractionation
of one charge by the other charges, so we naturally pick up the physics of
Phase II.

\listrefs

\end